\documentclass[%
 reprint,
superscriptaddress,
 amsmath,amssymb,
 aps,
floatfix,
]{revtex4-2}

\usepackage{booktabs}
\usepackage{svg}
\usepackage{float}
\usepackage{graphicx}
\usepackage{xcolor}
\usepackage{dcolumn}
\usepackage{bm}
\usepackage{soul}
\usepackage{tikz}
\usetikzlibrary{arrows.meta,positioning,calc,decorations.pathreplacing}
\usepackage[caption=false,labelformat=simple]{subfig}

\usepackage{graphicx}
\usepackage{physics}
\usepackage{siunitx}
\usepackage{amsthm}
\newtheoremstyle{query}%
{}{}
{\color{red}}
{}
{\sffamily\bfseries}{:}{12pt}
{}
\theoremstyle{query}
\newtheorem{aq}{Author Query/Comment}

\newcommand{\baq}{\begin{aq}}
\newcommand{\eaq}{\end{aq}}

\usepackage{hyperref}

\begin{document}

\title{Visible Spectral-Domain Optical Coherence Tomography for Photonic Integrated Circuits Characterization} 

\author{Yin Min Goh}
\affiliation{Research Laboratory of Electronics, Massachusetts Institute of Technology, Cambridge, MA 02139, USA}
\affiliation{These authors contributed equally to this work.}

\author{Chao Li}
\email[Corresponding author: ]{lichao@mit.edu}
\affiliation{Research Laboratory of Electronics, Massachusetts Institute of Technology, Cambridge, MA 02139, USA}
\affiliation{These authors contributed equally to this work.}

\author{Yunchan Hwang}
\affiliation{Research Laboratory of Electronics, Massachusetts Institute of Technology,
  Cambridge, MA 02139, USA}

\author{Helaman Flores}
\affiliation{Research Laboratory of Electronics, Massachusetts Institute of Technology,
  Cambridge, MA 02139, USA}

\author{\\ Mahmoud Jalali Mehrabad}
\affiliation{Research Laboratory of Electronics, Massachusetts Institute of Technology,
  Cambridge, MA 02139, USA}

\author{James G. Fujimoto}
\affiliation{Research Laboratory of Electronics, Massachusetts Institute of Technology,
  Cambridge, MA 02139, USA}

\author{Dirk R. Englund}
\affiliation{Research Laboratory of Electronics, Massachusetts Institute of Technology,
  Cambridge, MA 02139, USA}

\date{\today}

\begin{abstract}
Visible photonic integrated circuits underpin applications ranging from AR/VR to quantum control, yet lack a high-resolution, nondestructive diagnostic comparable to the optical frequency-domain reflectometry used in infrared silicon photonics. Here we adapt spectral-domain optical coherence tomography to measure guided-mode back-reflections in visible PICs. Broadband visible light injected into a circuit generates back-reflections that interfere with a depth-referencing local oscillator, and the resulting spectral fringes are recorded on a spectrometer. We validate the approach by resolving multiple round-trip echoes in a waveguide-coupled ring resonator using only single-port access. We then extend it to circuits integrated with diamond quantum micro-chiplets, clearly resolving input and output facets as well as PIC--QMC transition regions. The system achieves shot-noise-limited sensitivity, 50~dB dynamic range, 8~\(\mu\)m axial resolution in silicon nitride and a 2~mm imaging depth at 6~dB roll-off. SD-OCT therefore provides a practical, high-resolution diagnostic for visible PICs that uses a broadband probe source and requires only single-port optical access, enabling rapid characterization of propagation loss, backscattering and dispersion.
\end{abstract}

\maketitle

\section{Introduction}

Photonic integrated circuits (PICs) are rapidly becoming foundational to modern information technologies. 
At telecom wavelengths, PICs underpin low-latency interconnects for data centres and AI accelerators \cite{shen2017deep,hua2025integrated,ahmed2025universal}. 
At visible--near-infrared wavelengths, PICs enable compact and power-efficient beam steering for AR/VR and LiDAR \cite{shi2025flat,li2023frequency,lihachev2022low,zhang2022large, liang2024evolution}. They also provide scalable, stable, low-noise interfaces for quantum control and support highly multiplexed, label-free biosensing as well as lab-on-chip diagnostics \cite{psiquantum2025manufacturable,rogers2021universal,larocque2024tunable,luan2018silicon}. Operating in the VIS--NIR further leverages the mature silicon image-sensor and photodetector ecosystem, enables access to qubit platforms such as neutral atoms, trapped ions and diamond colour centres, and permits smaller device footprints owing to the shorter wavelength \cite{bigas2006review,poon2024silicon,christen2025integrated,li2024heterogeneous,govdeli2024room,chauhan2021visible,lin2022monolithically,lin2025optical,hogle2023high}. 
Mature low-loss dielectric platforms, including SiN and related high-bandgap materials, now support visible-wavelength routing, filtering, modulation and on-chip coupling through wafer-scale fabrication processes \cite{dong2022high,stanfield2019cmos,lin2022monolithically}, while multi-project wafer runs are beginning to deliver large heterogeneous systems rather than isolated test chips \cite{smith2023sin,stanfield2019cmos,lin2022monolithically,mahmudlu2023fully}. As visible PICs grow in scale and complexity—from centimetre-scale waveguide networks to multilayer hybrids incorporating MEMS or NEMS actuators—the bottleneck is no longer only design or fabrication \cite{sharif2023microcantilever,chen2024implantable,saha2026nanophotonic,saha2024photonic,qvotrup2024curved}. A central challenge is instead diagnostics: tools that can rapidly and non-destructively reveal how light propagates, reflects and disperses within the fabricated circuit \cite{che2025nanophotonic,morichetti2010roughness,shi2025hyperfine}.

Today’s methods fall short of that need. Laboratory verification still leans on bulky, slow setups: precision 3D stages to align lensed fibers to each port; cut-back or transmission measurements that collapse spatial information into a single number; and top-view camera inspection that is constrained by field of view, depth of focus and strong layer-stack scattering \cite{chanana2022ultra,he2019low,saha2026nanophotonic,starling2023fully}. Even in industrial settings with automated wafer probers, ``design-for-test'' rules---extra ports, monitor taps and fibre-array pitch constraints---consume area, perturb the circuit and limit what can be measured \cite{chrostowski2019silicon,hung2024design}. Crucially, transmission-based measurements rarely localize the origin of loss or reflection: they generally cannot distinguish distributed waveguide effects, such as sidewall-roughness scattering and bend radiation, from localized contributions at couplers, junctions, crossings or other discrete interfaces. In the infrared, optical frequency-domain reflectometry (OFDR) addresses this by mapping reflection versus distance using wide, mode-hop–free laser sweeps \cite{lee2012ultra,lai2017fiber,belt2017ultra,riemensberger2022photonic}. However, comparable tunable sources are scarce in the visible. What is missing for visible PICs is a platform-agnostic, fast, non-destructive, in-situ diagnostic that recovers guided-mode propagation loss, distributed and discrete back-scattering, and dispersion with micrometre-scale resolution, millimetre-scale imaging depth and high dynamic range---without redesigning the chip around test fixtures, and without sacrificing compatibility with complex 3D photonic, MEMS/NEMS and hybrid quantum architectures.

\begin{figure*}[htbp]         \includegraphics[width=0.98\textwidth]{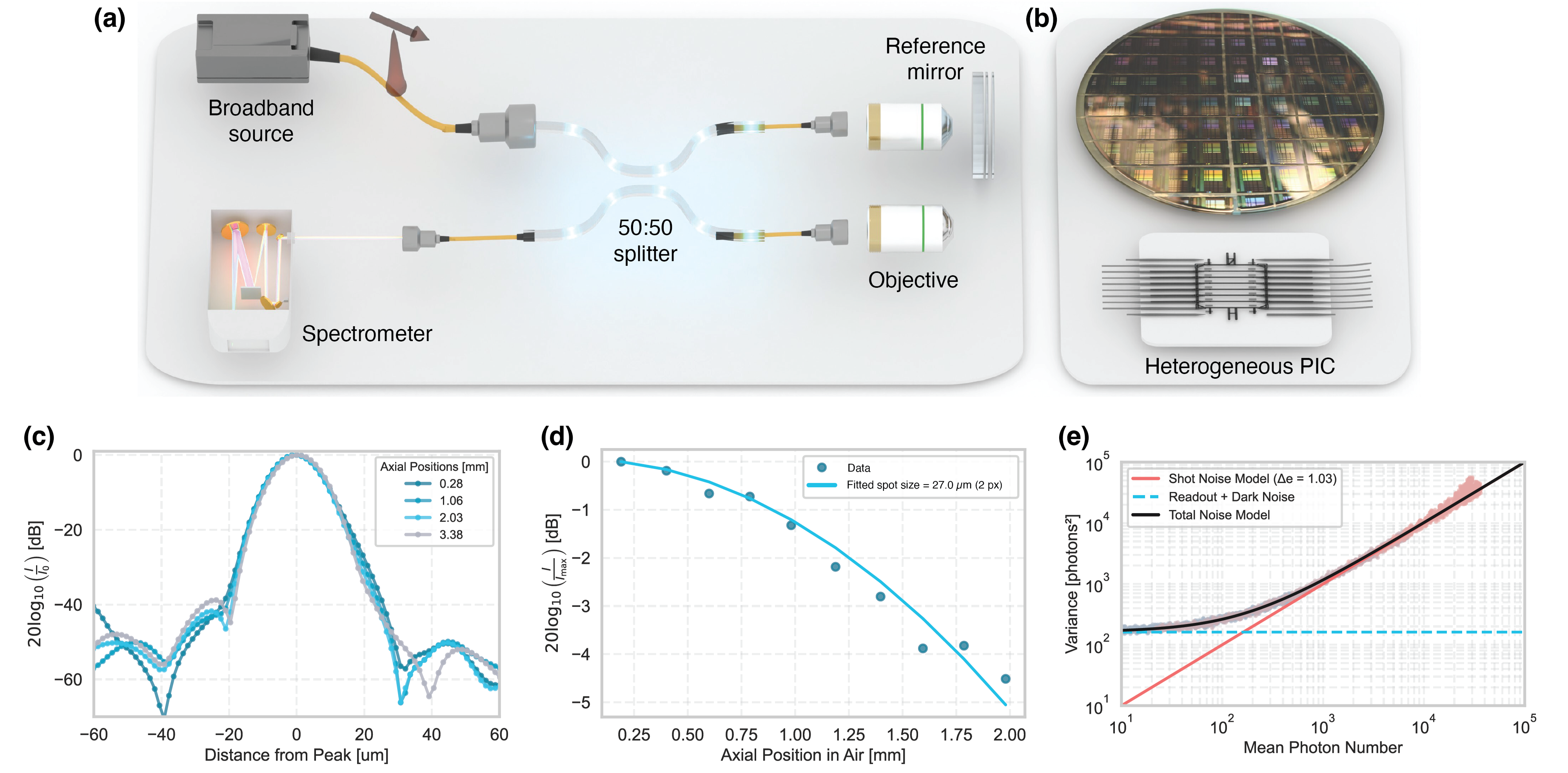}
         \caption{Overview of OCT for PICs inspection. (a) CAD rendering of the optical setup: a fibre-based Michelson interferometer delivers the combined sample--reference field to a commercial spectrometer. Interference with the reference arm amplifies weak backscattering from PICs---often treated as a nuisance---thereby revealing circuit loss, dispersion and coupling losses across architectures that are otherwise difficult to diagnose. (b) Representative inspection contexts and length scales for visible-PIC OCT: a 4-inch wafer containing many millimetre- to centimetre-scale devices (top), and a heterogeneously integrated QMC-on-PIC socket (bottom). In practice, the OCT probe in (a) can be scanned across intact wafers or diced chips before or after dicing. The shown diamond QMC on PIC is about \(\sim 100\,\mu\mathrm{m}\times 30\,\mu\mathrm{m}\). (c) Calibration of axial resolution and dynamic range. (d) Imaging depth and spectral roll-off. (e) Detection sensitivity with shot-noise-limited performance per CCD pixel.}
         \label{fig:overview}
\end{figure*}

We close this gap by adapting spectral-domain optical coherence tomography (SD-OCT) to visible--NIR PIC diagnostics. OCT, widely used in biomedical and industrial metrology, measures depth-resolved back-reflections via low-coherence interferometry \cite{huang1991optical,drexler2008optical}. Here we repurpose it as a single-port reflectometric tool for on-chip guided light. A broadband source illuminates the circuit through a standard input coupler or cleaved edge; the weak field that back-scatters or reflects from imperfections and interfaces returns to the same port and interferes with a fixed-delay reference. Instead of wavelength sweeping, a spectrometer records the interference spectrum in one shot; Fourier transfor yields an axial ``A-scan'' of reflectivity versus position along the optical path, with axial resolution set by the source bandwidth and imaging range set by spectral sampling. 

In our implementation, broad spectral width and high spectral sampling deliver $\approx 16\,\mu\mathrm{m}$ axial resolution (in air) and $\ge 2\,\mathrm{mm}$ imaging depth per acquisition window, sufficient to localize sub-component features along millimetre-scale propagation lengths in the PIC. Translating the reference arm stitches windows to cover centimetre-depth circuits and beyond. A dispersion-aware reconstruction restores the point-spread function, while model-based inversion---informed by the PIC layout---recovers per-element loss and reflectance with shot-noise-limited sensitivity. Since the method couples only at a standard input and reads out the same port, it eliminates design-for-test overhead, can operate across diverse material platforms, and natively accommodates 3D and hybrid assemblies (e.g., MEMS-tunable photonics and PIC--quantum-memory composites). Beyond manufacturing yield and debugging, depth-resolved backscatter maps enable device physics studies---for example, identifying reflection-induced noise in lasers and amplifiers, or testing claims of backscattering suppression in topological waveguides \cite{rosiek2023observation}. By bringing OCT’s depth sectioning and sensitivity to guided-wave photonics in the visible, this work demonstrates a practical metrology blueprint that scales with PIC complexity and helps designers optimise what matters: how light actually flows on the chip.

\section{Results}

\begin{figure*}[htbp]
    \includegraphics[width=0.98\textwidth]{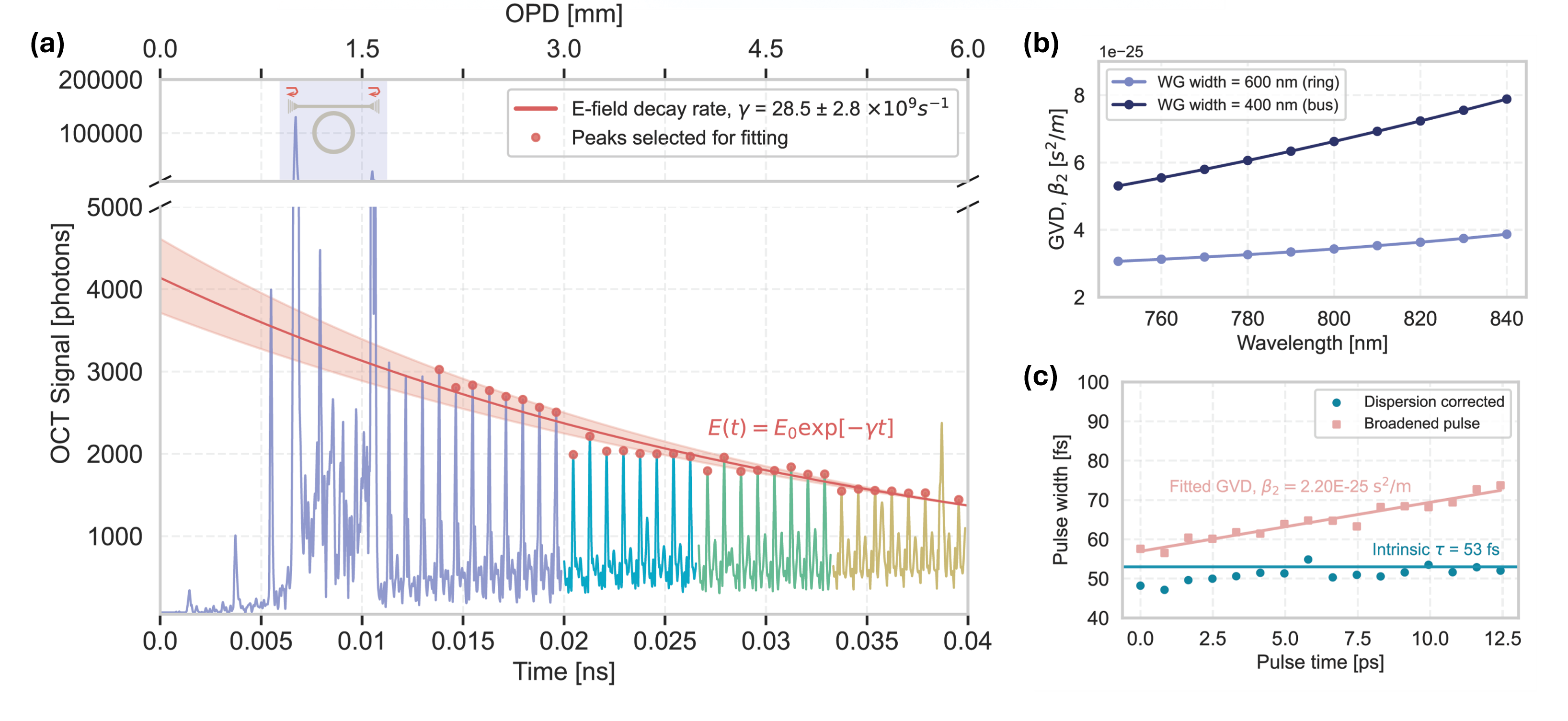}
         \caption{
         (a) OCT reconstruction reveals that a short input pulse yields through-port echoes spaced by the round-trip time $T_{\mathrm{rt}}$; a grating-coupler back-reflection re-excites the ring on the return path, producing a delayed echo train. The detected amplitude decay captures the total loss in the ring resonator.
        (b) FDTD simulations of two simple waveguides with different widths.
        (c) The red curve shows echo broadening due to group-velocity dispersion. The cyan curve shows restoration of the broadened pulse to its intrinsic pulse width, $\tau = 53\,\mathrm{fs}$ (equivalent to $16\,\mu\mathrm{m}$ axial resolution), using a two-step numerical dispersion-compensation procedure. The model derivation is provided in the SI.
        }
         \label{fig:ring}
\end{figure*}

\subsection{Setup and experimental workflow}

Fig.~\ref{fig:overview} depicts the spectral-domain optical coherence tomography (SD-OCT) setup used to probe guided-mode back-reflections from PICs. The system is a modified Michelson interferometer with the sample and reference arms matched in optical path length. Unlike clinical OCT used in ophthalmology, the sample arm employs microscope objectives to efficiently couple light into the tightly confined waveguide modes at visible wavelengths (e.g., $100\times$, NA 0.95 for cleaved edge couplers; $10\times$, NA 0.25 for grating couplers). The reference arm replicates the sample arm optics to balance dispersion, but replaces the device under test with a metallic mirror positioned at the focal plane to form a cat’s-eye retroreflector \cite{thompson2012narrow}, improving stability and alignment tolerance for the return field. A 50:50 fibre splitter serves as the interferometer beamsplitter. Off-axis parabolic collimators launch and collect the beams in both arms, eliminating chromatic aberration and minimising spurious OCT artefacts from surface Fresnel reflections by reducing transmissive optics. Free-space polarisers define the TE input mode relative to the PIC, and in-line polarisation controllers match the returning linear polarisation states in the two fibre arms to maximise fringe visibility.

Weak back-reflections from the sample arm interfere coherently with the reference field, effectively amplifying the signal through field multiplication and enabling detection with shot-noise-limited sensitivity. A high-resolution spectrometer records the resulting interferogram $I(k)$, which encodes depth-dependent backscatter as spectral modulations. While the broadband source sets the axial resolution via its coherence length ($\sim \lambda_c^2/\Delta\lambda$), the imaging depth ($\sim \lambda_c^2/\delta\lambda$) is governed by how finely the spectrometer samples the spectrum due to the Nyquist limit. By demultiplexing the broadband light into narrow spectral bins, each pixel captures a small $\delta\lambda$, allowing interference from deeper reflectors to be resolved. Following wavelength-to-$k$ calibration and dispersion correction, a Fourier transform of $I(k)$ yields an axial reflectivity profile (A-scan), revealing the locations and amplitudes of distributed scattering and discrete reflections along the waveguide path.

\begin{figure*}[htbp]
         \includegraphics[width=0.95\textwidth]{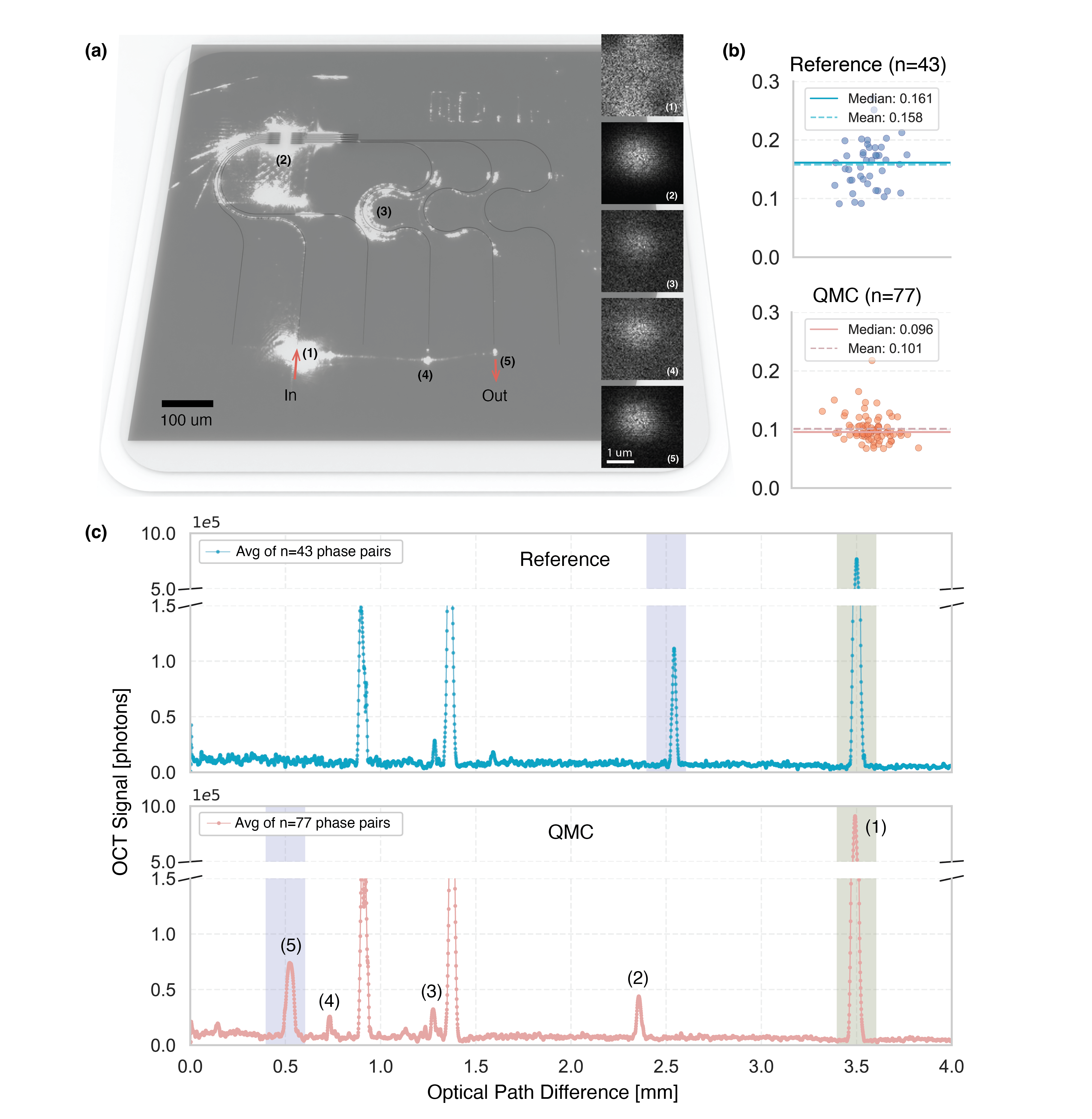}
         \caption{(a) Top-view scattering image of the PIC, with five B-scan profiles on the right corresponding to the labeled depth locations within the circuit. (b) Ratio of back-reflection strengths from the input and output facets for a QMC channel compared with a reference loop. (c) The averaged roll-off corrected A-scan profiles of the QMC channel and the reference loop.}
         \label{fig:hybridcircuit}
\end{figure*}

\subsection{Technique Validation via a Bus Waveguide Coupled to a Ring Resonator}

To validate the method, we apply it to a single-mode bus waveguide slightly over-coupled to a $\mathrm{Si_3N_4}$ ring resonator (Fig.~\ref{fig:ring}). Broadband light from a superluminescent diode ($\lambda_c = 800\,\mathrm{nm}$, $\Delta\lambda = 30\,\mathrm{nm}$) is launched into the bus. The broadband field can be viewed as a short temporal wave packet that is partitioned by the bus–ring coupler into a directly transmitted component and a sequence of delayed replicas arising from successive round trips in the ring.

In the single-port backscattering configuration, the output grating provides a weak reflection that redirects this entire sequence back through the device. On the return pass, the reflected components again interact with the coupler, producing additional delayed contributions. The measured reflectogram therefore consists of peaks associated with specific forward–backward propagation pathways whose total delays coincide \cite{gottesman2004new}. Peak spacing directly yields the ring optical path length, while the systematic amplitude evolution reflects coherent addition of pathways of identical total round-trip order. Because this structure is governed solely by the coupling coefficient and the intrinsic round-trip loss, both parameters can be extracted quantitatively from a single broadband measurement.

Since the single-shot imaging depth is limited by the spectrometer, we incrementally scan the reference arm and acquire more than 30 echoes over a total range of $4.5~\mathrm{mm}$. The resulting segments are numerically stitched to reconstruct a continuous depth profile, as described in the SI. After dispersion compensation, the fitted envelope (Fig.~\ref{fig:ring}) directly reveals the resonator decay dynamics. Since the detected OCT signal is proportional to the interference term $E_{\mathrm{sig}} E_{\mathrm{lo}}$, and the local-oscillator field $E_{\mathrm{lo}}$ is constant in time, the measured amplitude faithfully traces the temporal decay of the backscattered field $E_{\mathrm{sig}}(t)$. Fitting the echo-train envelope with $E_{\mathrm{sig}}(t) \propto \exp(-t/\tau_r)$ yields a decay constant of $\tau_r \approx 0.038~\mathrm{ns}$. This corresponds to a loaded quality factor of $Q=\omega_0\tau_{\mathrm{energy}} \approx 4.5\times10^{4}$, where $\omega_0$ is the resonance frequency and $\tau_{\mathrm{energy}}=\tau_r/2$ is the resonator energy-decay time. The extracted $Q$ agrees to within $10\%$ with independent transmission-based measurements (see SI), which indicate an intrinsic quality factor of $Q_i \approx 1.8\times10^{5}$ and a propagation loss of $4.0~\mathrm{dB\,cm^{-1}}$ \cite{stanfield2019cmos}. These results validate our OCT-based approach while eliminating the need for access to an output port.

\subsection{Diagnostics for Hybrid Quantum Photonics}

We next applied this technique to diagnose a PIC incorporating on-chip polarizers, beam splitters, cleaved couplers and a socket interface to diamond waveguide arrays hosting quantum memories~\cite{chen2024scalable,starling2023fully}. These hybrid PICs are designed to address the emission lines of diamond colour centres, near \(\sim 737\,\mathrm{nm}\) for silicon vacancies and \(\sim 620\,\mathrm{nm}\) for tin vacancies. To probe this spectral range, we configured a broadly tunable supercontinuum source to deliver a 40-nm bandwidth centred at \(700\,\mathrm{nm}\) for the measurements reported below.

Because the guided-mode field diameter is sub-micrometre, we first performed a transverse raster scan over a targeted \(3\,\mu\mathrm{m} \times 3\,\mu\mathrm{m}\) area to localize the guided mode. At each of \(75 \times 75\) uniformly spaced \((x,y)\) positions, we recorded 20 A-scans and applied the phase-ramp technique to each A-scan. 
The resulting map reveals multiple two-dimensional Gaussian features arising from index-mismatch-induced reflections: (2) from the diamond QMC, (3) from the polarizer perturbation, and (4)–(5) from the output facets of the PIC. Their axial separations agree with the PIC GDS layout. Residual spurious features are attributed to the measurement setup because they shift oppositely to the PIC features as we moved the retroreflector, enabling unambiguous identification of the waveguide signal for subsequent analysis (see Fig.~\ref{fig:hybridcircuit}(c)).

From the relative amplitudes of the facet back-reflections at locations (4) and (5), with OPDs of 0.75\,mm and 0.53\,mm, respectively, we extracted a splitting ratio of \(\sim 1{:}3\), in good agreement with the design, in which the strongly coupled channel is used to collect emitted photons and the weakly coupled channel is used for resonant excitation \cite{parker2024diamond}.

We also measured a short reference loopback channel. Given their short lengths, we neglect propagation loss in both the reference and QMC channels and compare the two datasets directly. The output-facet echo of the QMC channel is weaker than that of the reference channel because the guided light experiences both the QMC--PIC transition loss and the splitter loss. Using the measured splitting ratio, we infer a QMC--PIC transmission coefficient of \(0.8\), corresponding to a \(-0.8\)\,dB loss. Detailed calculations for extracting the splitting ratio and the diamond-to-PIC loss are provided in the SI.

\section{Discussion}

This work addresses key challenges in diagnosing PICs with increasingly complex architectures—including three-dimensional interconnects and operation in the visible regime—by adapting spectral-domain OCT into a single-port, depth-resolved reflectometry platform for guided modes \cite{saha2026nanophotonic,saha2024photonic,danielson1987guided,takada1987new}. By leveraging interferometric gain from a stable reference field and combining it with dispersion-aware reconstruction and layout-informed inversion, the method localizes and quantifies distributed backscatters and discrete reflections over cm-scale optical paths with $\sim 8\,\mu\mathrm{m}$ axial resolution, without requiring output access, monitor taps, or ``design-for-test'' overhead \cite{chrostowski2015silicon}. 

Validation on ring resonators shows that time-of-flight ``echo trains'' in the reflectogram are not merely qualitative signatures, but encode coupling and intrinsic loss in a manner that enables direct parameter extraction from single-port measurements. The hybrid quantum-photonics case study further demonstrates that the same depth sectioning isolates specific interfaces (splitters, facets and the QMC--PIC transition), allowing component-level losses to be quantified \emph{in situ}.

Beyond component-level debugging, depth-resolved OCT reflectograms turn otherwise “nuisance” backscatter into a spatially localized metric of where disorder-induced reflections build up along the guided path—information that is largely averaged out in transmission. This is particularly consequential for resonator-rich PICs because coherent backscattering in microring resonators is cavity-enhanced and increases quadratically with the effective group index, making sidewall roughness a first-order constraint for high-$Q$ and slow-light operation \cite{morichetti2010coherent}. Although we do not study topological devices here, the same localization capability is directly relevant to emerging visible ring-resonator lattices (e.g., near-visible SiN ring arrays for topological photonics), where performance can hinge on how scattering is distributed across many coupled sites rather than on a single lumped loss number \cite{sharp2024near}. 

Looking ahead, established OCT methodologies from ophthalmology could further expand the diagnostic toolbox for PIC characterization. Split-spectrum (sub-band) reconstruction—widely adopted in OCT angiography—forms images from discrete spectral portions, intentionally trading axial resolution for enhanced robustness and spectroscopic contrast \cite{jia2012split}. In the visible regime, related spectroscopic OCT approaches have enabled functional contrasts such as retinal oximetry and depth-resolved spectral shifts (redshifts) associated with wavelength-dependent scattering \cite{pi2020retinal,meng2025imaging}. Spectral shaping strategies have also been demonstrated \cite{zhang2019improving}, indicating that adaptive control of the source spectrum can dynamically improve SNR and enhance feature extraction.

Translating these concepts to PIC reflectometry would yield depth-resolved spectra of backscatter and attenuation along guided modes, enabling wavelength-dependent loss to be assigned to specific circuit elements. Such spectroscopic sectioning could help disentangle distributed roughness-limited propagation from localized defects at junctions or couplers, and from material absorption—thereby extending OCT from structural inspection to functional, wavelength-aware diagnostics of integrated photonic circuits.

\bibliography{references}

\section{Methods }

\subsection{Additional implementation details of the SD-OCT setup}

The interferometer output was routed through 20~m of fibre to a high-resolution spectrometer located in a separate laboratory. The spectrometer uses an all-reflective layout with a 1200~lines/mm diffraction grating and matched toroidal mirrors for collimation and focusing (both \(f=\SI{315}{\milli\meter}\)). The fibre tip is placed at the collimator focal plane, \SI{315}{\milli\meter} from the first mirror. The collimated beam is dispersed by the grating and re-imaged onto a CCD at the focal plane of the second mirror. We use a PIXIS 2K camera (\(2048\times512\), \SI{13.5}{\micro\meter} pixel pitch). The CCD operating temperature ranged from approximately \(5\,^\circ\mathrm{C}\) to \(-25\,^\circ\mathrm{C}\) owing to aging of the CCD TEC and failure of its cooling fan, which in turn led to increased readout noise and dark current. For narrow-linewidth sources, the optics act as a relay, mapping the fibre mode directly onto the detector; for a monochromatic input, we measured that the imaged fibre mode spans approximately one to two pixels.

Phase-stepped interferometry was implemented to isolate the sample--reference cross-correlation term from the DC background and sample autocorrelation, and to suppress the complex-conjugate ambiguity about the reference delay \(z_{\mathrm{ref}}\). A piezoelectric actuator in the reference arm introduced calibrated phase offsets between consecutive acquisitions.

\subsection{Dispersion Compensation}
To accurately interpret both the amplitude and width of these echoes, we next restore the intrinsic axial resolution of the spectral-domain OCT system by numerically compensating dispersion. We implement a two-step routine. First, we apply a second-order spectral phase term (Supplementary Section~SX) chosen to maximize the sharpness of the earliest reflection from the input grating coupler. This step corrects the dominant dispersion mismatch introduced by the \(\approx 1~\mathrm{m}\) delivery fibers in both arms, the fused fiber splitter, and bulk optics, returning the point-spread function to the transform limit set by the \(30~\mathrm{nm}\)-bandwidth SLD.

In a second step, we remove the residual group-velocity dispersion (GVD) of the \(\mathrm{Si_3N_4}\) waveguide itself, informed by finite-difference time-domain (FDTD) simulations of the guided mode. We apply an additional quadratic phase adjustment that minimizes the dependence of the echo width on propagation distance, thereby compensating waveguide-induced pulse broadening. This correction narrows the late-echo full width at half maximum (FWHM) by \(20\)--\(30~\mathrm{fs}\), in good agreement with the broadening predicted from the simulated waveguide GVD. From the linear dependence of the residual dispersion on propagation length (Fig.~\ref{fig:ring}(c)), we extract a GVD of \(\beta_2 \approx 2.2\times 10^{-25}~\mathrm{s}^2/\mathrm{m}\), which is consistent with the FDTD value of \(3.4 \times 10^{-25}~\mathrm{s}^2/\mathrm{m}\) for this waveguide geometry.


\subsection{Camera Inspection}
Camera images were acquired using a Thorlabs CS165MU positioned above the sample to monitor coupling from the chip edge and the subsequent propagation of the broadband light within the PIC. The SuperK light source was set to a center wavelength of 700\,nm with a spectral width of 40\,nm, and the input polarization was set to horizontal.
 The camera exposure time was 100\,ms with a gain of 20\,dB. The input optical power was approximately 40\,\textmu W. Scattered light from the PIC was collected using an objective with an approximate focal length of 2\,cm. Together with three additional 1\,inch lenses used for optical relay and imaging, the top-view imaging system had an overall magnification of approximately 4.5.

\subsection{FDTD Simulations}
The SiN bus waveguide was \(300~\mathrm{nm}\) thick and \(375~\mathrm{nm}\) wide, supporting a single TE mode in the \(700\text{--}800~\mathrm{nm}\) wavelength range. The cladding was \(\mathrm{SiO_2}\). The ring resonator waveguide had the same thickness and a width of \(600~\mathrm{nm}\). The outer diameter of the ring resonator was \(40~\mu\mathrm{m}\), and the gap between the ring and the bus waveguide was \(300~\mathrm{nm}\). We performed FDTD simulations in Tidy3d to extract the dispersion of the our waveguides of different widths vs. wavelengths.

\section{Data availability}
The data that support the plots and findings within this paper are available from the corresponding author on request.

\section{Author contributions}
CRediT author statement --
Conceptualization: DRE, CL.
Methodology: YMG, CL, YH, JGF.
Software: YMG, CL, YH.
Validation: CL, YMG.
Formal analysis: YMG, CL, YH, MJM, HF. 
Investigation: YMG, CL.
Resources: DRE, JGF, CL.
Data curation: YMG, CL.
Writing – original draft: CL. 
Writing – review \& editing: CL, YMG, YH, MJM, HF.
Visualization: YMG, CL, YH, MJM, HF.
Supervision: CL, DRE.
Project administration: CL, MJM.
Funding acquisition: DRE, JGF, CL.

\section{Competing interests}
DRE is CSO and co-founder of Axiomatic\_AI. The other authors declare no competing interests.

\section{Acknowledgment}
We thank Drs.\ Sofia Patomäki, David J.~Starling, P.~Benjamin Dixon, Adrian Messen, Gen Clark and Matt Saha for providing test devices and red team reviews, Avinash Kumar for lending us a free-space Michelson interferometer for initial tests, Saumil Bandyopadhyay and Juejun Hu for helpful discussions. The authors acknowledge the support of Flexcompute for providing free Tidy3D credits used in this work.

\section{Funding}
This work was supported by the NSF Engineering Research Center for Quantum Networks (grant no. EEC-1941583), CQN (EEC-1941583). The authors also acknowledge partial support from the Quantum Moonshot Program. DRE, YMG and CL acknowledge partial support from the Quantum Moonshot Program.


\end{document}